\documentclass[reprint,
superscriptaddress,
nofootinbib,
 amsmath,amssymb,
 aps,
prl,
]{revtex4-2}
\usepackage{xcolor}
\definecolor{darkblue}{rgb}{0,0,.6}
\usepackage[colorlinks=true, citecolor=darkblue, linkcolor=darkblue, menucolor=darkblue, urlcolor=darkblue]{hyperref}
\usepackage{float}

\usepackage{graphicx}
\usepackage{physics}
\usepackage{dcolumn}
\usepackage{bm}
\usepackage{multirow}
\usepackage{amsmath}
\usepackage{array}
\newcommand{\AMRO}{$\mathrm{A_2MgReO_6\ 
(A=Ca,Sr,Ba)}$}
\newcommand{\AMROx}{$\mathrm{A_2MgReO_6}$}
\newcommand{\BMRO}{$\mathrm{Ba_2MgReO_6}$}
\newcommand{\SMRO}{$\mathrm{Sr_2MgReO_6}$}
\newcommand{\CMRO}{$\mathrm{Ca_2MgReO_6}$}

\newcommand{\RE}{$\mathrm{Re}$}

\newcommand{\Dii}{$\mathrm{\Delta_2}$}
\newcommand{\Diii}{$\mathrm{\Delta_3}$}
\newcommand{\OX}{$\mathrm{O}$}
\newcommand{\REO}{$\mathrm{ReO_6}$}
\newcommand{\jes}{$j_{eff}=1/2$}
\newcommand{\jgs}{$j_{eff}=3/2$}
\newcolumntype{C}[1]{>{\centering\let\newline\\\arraybackslash\hspace{0pt}}m{#1}}

\begin{document}

\preprint{APS/123-QED}

\title{Spin-orbit-lattice entangled state in \texorpdfstring{\AMRO{}}{A\_2MgReO\_6} revealed by resonant inelastic X-ray scattering}

\author{Felix I. Frontini}
 \email{felix.frontini@mail.utoronto.ca}
 \affiliation{Physics Department, University of Toronto, 60 St. George Street, Toronto, ON, M5S 1A7}
\author{Graham H.J. Johnstone}
 \affiliation{Physics Department, University of Toronto, 60 St. George Street, Toronto, ON, M5S 1A7}
\author{Naoya Iwahara}
 \affiliation{Graduate School of Engineering, Chiba University, 1-33 Yayoi-cho, Inage-ku, Chiba-shi, Chiba 263-8522, Japan} 
\author{Pritam Bhattacharyya}
 \affiliation{Institute for Theoretical Solid State Physics, Leibniz IFW Dresden, Helmholtzstra{\ss}e~20, 01069 Dresden, Germany}
\author{Nikolay A. Bogdanov}
 \affiliation{Max Planck Institute for Solid State Research, Heisenbergstra{\ss}e~1, 70569 Stuttgart, Germany}
\author{Liviu Hozoi}
 \affiliation{Institute for Theoretical Solid State Physics, Leibniz IFW Dresden, Helmholtzstra{\ss}e~20, 01069 Dresden, Germany}
\author{Mary H. Upton}
 \affiliation{Advanced Photon Source, Argonne National Laboratory, 9700 S. Cass Avenue, Lemont, IL 60439}
\author{Diego M. Casa}
 \affiliation{Advanced Photon Source, Argonne National Laboratory, 9700 S. Cass Avenue, Lemont, IL 60439}
\author{Daigorou Hirai}
 \affiliation{Department of Materials, Physics and Energy Engineering, Nagoya University, Furo-cho, Chikusa-ku, Nagoya, 464-8601, Japan}
\author{Young-June Kim}%
 \email{youngjune.kim@utoronto.ca}
 \affiliation{Physics Department, University of Toronto, 60 St. George Street, Toronto, ON, M5S 1A7}%

\date{\today}

\begin{abstract}
The $5d^1$ ordered double perovskites present an exotic playground for studying novel multipolar physics due to large spin-orbit coupling.
We present Re $L_3$ edge resonant inelastic X-ray scattering (RIXS) results that reveal the presence of the dynamic Jahn-Teller effect in the \AMRO{} family of $5d^1$ double perovskites. 
The spin-orbit excitations in these materials show a strongly asymmetric lineshape and exhibit substantial temperature dependence, indicating that they are dressed with lattice vibrations.
Our experimental results are explained quantitatively through a RIXS calculation based on a spin-orbit-lattice entangled electronic ground state with the dynamic Jahn-Teller effect taken into consideration.
We find that the spin-orbit-lattice entangled state is robust against magnetic and structural phase transitions as well as against significant static Jahn-Teller distortions.
Our results illustrate the importance of including vibronic coupling for a complete description of the ground state physics of $5d^1$ double perovskites.
\begin{description}
\item[Usage]
Secondary publications and information retrieval purposes.
\end{description}
\end{abstract}

\maketitle


One of the major themes in quantum materials research in the past decade has been materials with large spin-orbit coupling (SOC) and its impact on their magnetic properties.
Although particular emphasis has been placed on the $5d^5$ physics of iridates, other $5d$ transition metal (TM) ion systems have been drawing interest in recent years, including the $5d^1$ ordered double perovskite (DP) systems  ($5d^1\mathrm{=\ W^{5+},\ Re^{6+},\ or\ Os^{7+}}$) \cite{5d1_chen,5d1_tungstenates,CMRO2003,5d1_rhenates1,BMRO2019,BMRO2020,SMRO2020,5d1_rhenates2,5d1_rhenates3,5d1_osmates1,5d1_osmates2,5d1_osmates3,5d1_osmates4}.
Crucially though, the typical atomic picture effective in understanding iridate physics does not satisfactorily explain the physics of the $5d^1$ DPs.
In the $5d^1$ DPs of formula $\mathrm{A_2BB'O_6}$ the sole magnetic $5d^1$ ion occupies the $\mathrm{B'}$ site and possesses a four-fold degenerate ground state configuration described by \jgs{}.
Configured as such, the net magnetic moment $M=2s-l$ vanishes as spin and orbital components of the angular momentum cancel each other exactly \cite{5d1_chen}.
The atomic description breaks down, however, in the case of real materials which broadly possess finite magnetic moments \cite{5d1_rhenates2,5d1_osmates1,CMRO2003}. 
The presence of a suppressed but non-zero magnetic dipole moment is typically attributed to hybridization between the spatially extended TM-$5d$ orbitals and the ligand \OX{}-$2p$ orbitals \cite{5d1_chen}.
However, coupling to the lattice degrees of freedom has also been shown theoretically to have a significant effect on the magnetic moment size in $5d^1$ systems \cite{5d1_chen,Liviu_vibronic,Naoya_vibronic}.
Moreover, vibronic coupling to Jahn-Teller (JT) active modes  (symmetric deformations of the TMO$_6$ octhedra) can also explain the puzzling suppression or complete lack of static JT distortions observed in many $5d^1$ DPs \cite{5d1_osmates2,5d1_cubic,Naoya_vibronic,Naoya_vibronic2}.
The dynamic JT effect creates an average JT distortion which lowers the ground state energy whilst relaxing the energetic advantage of static distortions, such that cubic structures are stabilized.\par
\begin{figure}[t]
    \centering
    \includegraphics[width=0.95\columnwidth]{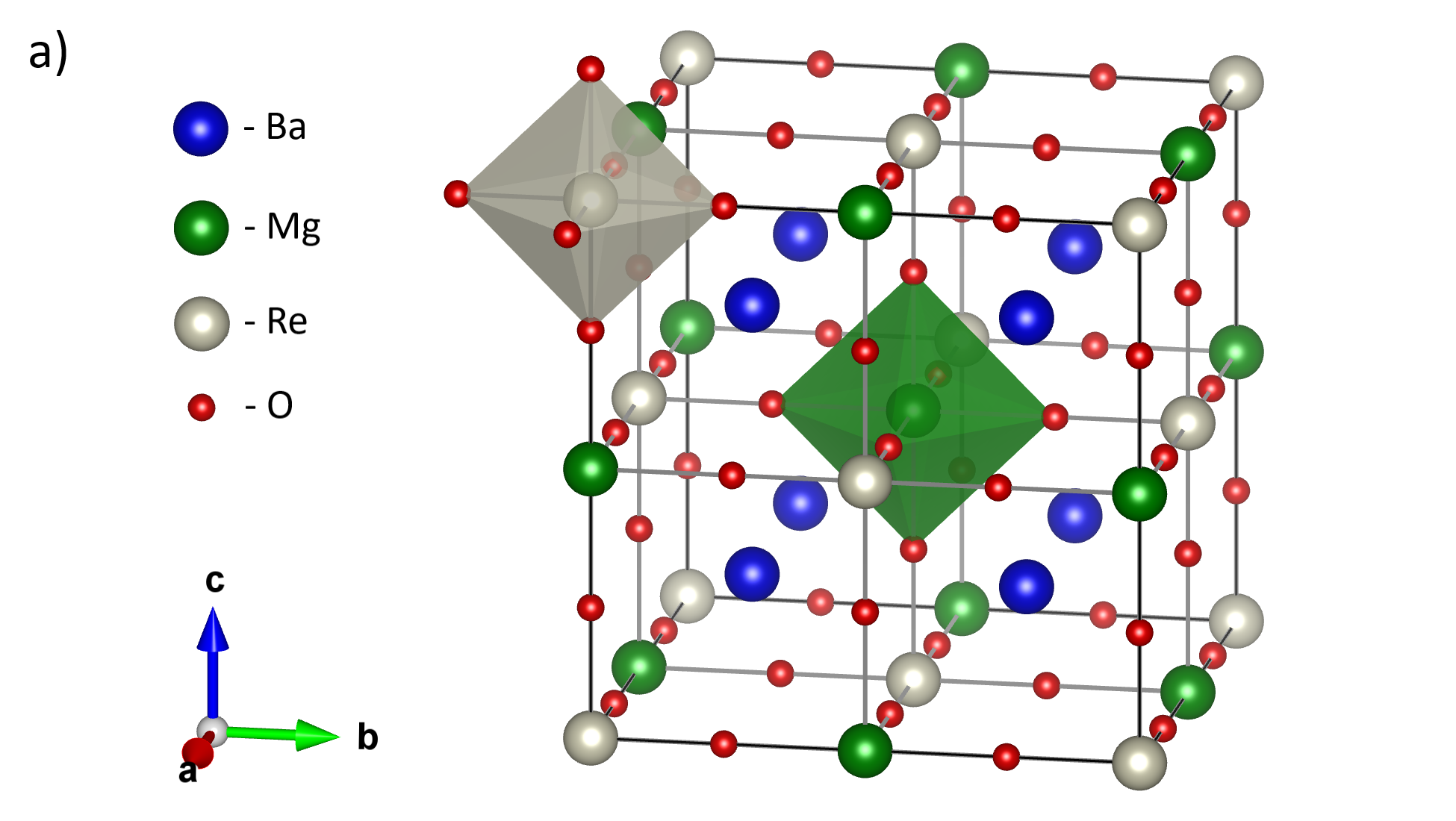}
    \includegraphics[width=0.95\columnwidth]{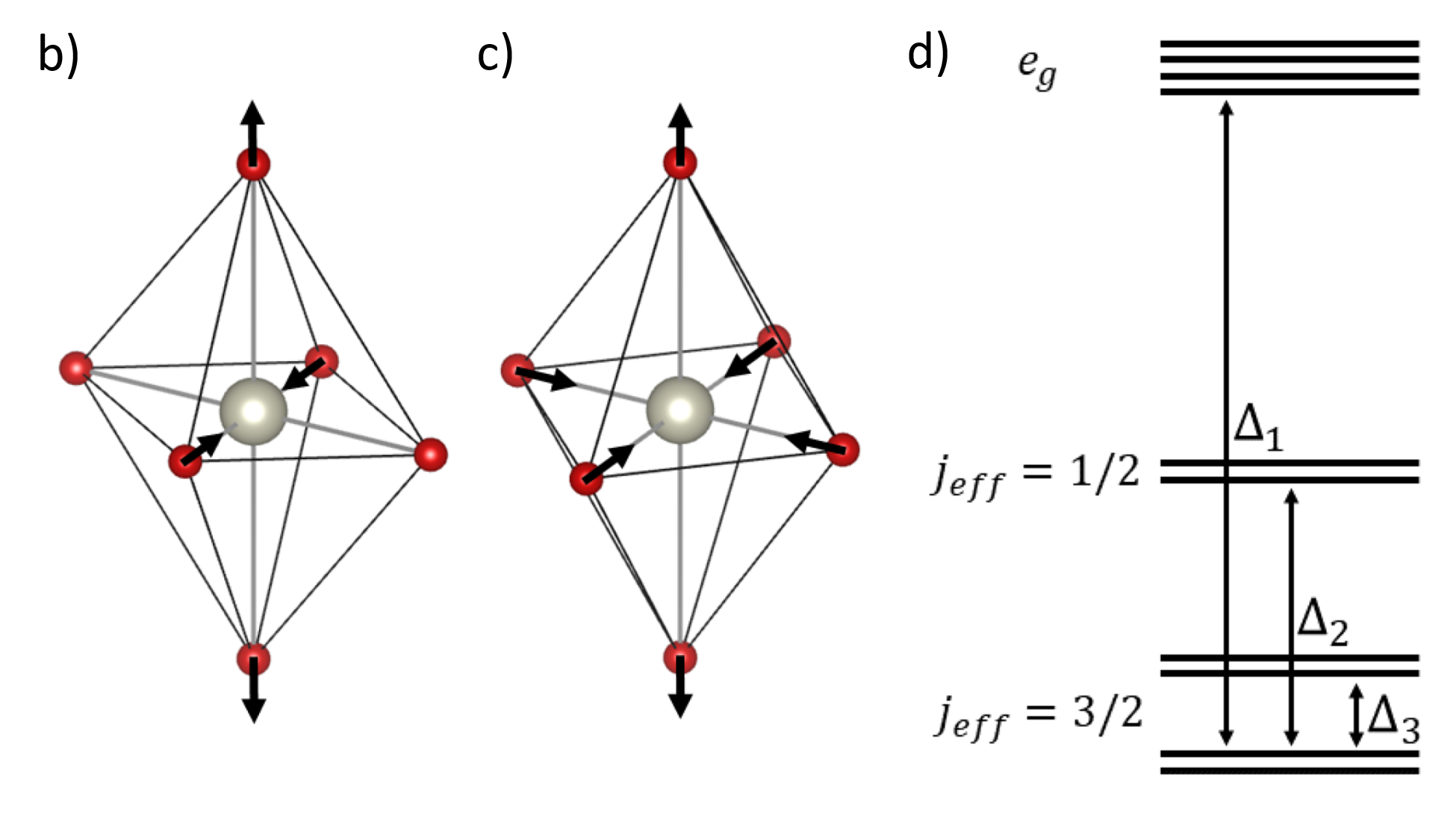}
    \caption{a) \BMRO{} undistorted cubic double perovskite crystal structure at 300 K (space group $Fm\Bar{3}m$). Schematic of \REO{} octahedron distortion in b) \CMRO{} and c) \SMRO{}. d) Energy splittings of the $5d$ levels in \AMROx{} due to crystal field, SOC, and static JT distortion effects. Note that $\Delta_3$ = 0 in the absence of static distortion (e.g. \BMRO{} above $T_q$).}
    \label{fig:fig1}
\end{figure}

The \AMRO{} systems allow us to study vibronic coupling in $5d^1$ DPs by systematically varying the A-site ions.
In particular, \BMRO{} constitutes a compelling canditate for the formation of a spin-orbit-lattice entangled state, in which the spin-orbit ground state is strongly modified by coupling to lattice degrees of freedom.
\BMRO{} crystallizes in an undistorted cubic DP structure at room temperature shown in Fig. \ref{fig:fig1} a) before undergoing a subtle structural transition to a tetragonal structure associated with the onset of quadrupolar order, marked by the slight cooperative distortion of \REO{} octahedra \cite{BMRO2020}.
Despite this deviation from the cubic ideal, the minimal structural distortion in \BMRO{} stands in stark contrast to the much larger distortion present in its structural analogues; tetragonal \SMRO{} and monoclinic \CMRO{}, in which the \REO{} octahedra show significant static distortion as illustrated in Fig. \ref{fig:fig1} b), c) \cite{SMRO2003,CMRO2003}.
The relative lack of static JT distortion in \BMRO{}, then, seems to indicate the presence of the dynamic JT effect.
Moreover, recent theoretical studies have shown that vibronic coupling could function as a mechanism for the formation of the anomalous ferroic component of the quadrupolar order in \BMRO{} \cite{Naoya_vibronic2}.
In addition to the structural clues, we note that \BMRO{} also fits the expectation for a vibronically coupled $5d^1$ DP as a magnetic material, ordering magnetically at $T_m$ =18 K with a reduced but non-zero magnetic moment of $\sim 0.7\ \mu_B$ \cite{BMRO2019}.
The structural analogues of \SMRO{} and \CMRO{} furthermore present an opportunity to investigate the effect of static JT distortions on the presence and strength of vibronic coupling in $5d^1$ DPs. \par
In this Letter, we present resonant inelastic X-ray scattering (RIXS) measurements of the \AMRO{} family of $5d^1$ double perovskites that reveal a spin-orbit-lattice entangled ground state in all members.
The presence of such a state not only in \BMRO{} but also in \SMRO{} and \CMRO{} furthermore suggests that vibronic coupling in $5d^1$ DPs is remarkably robust against the presence of static JT distortions.
The presence of a spin-orbit-lattice entangled ground state is indicated by careful analysis of the RIXS spectra and their temperature dependence.
We observe that, common to all three samples, the \jgs{} $\to 1/2$ spin-orbit excitation displays a significant asymmetry indicative of dressing by lattice vibrations.
The excitation is also observed to broaden and shift to higher energy with temperature, suggesting a thermally activated effect in keeping with a phonon dressing origin.
A microscopic model is constructed to simulate a vibronically coupled RIXS spectrum and shows good agreement with our data.
Our study thus reveals the presence of the dynamic JT effect in \AMRO{} and lends credence to the assertion that the dynamic JT effect stabilizes the observed multipolar physics in \BMRO{}.\par
Our experiments were performed with single crystal samples of \BMRO{} and \SMRO{} as well as a powder sample of \CMRO{}. 
The growth of \BMRO{} and \SMRO{} have been previously reported \cite{BMRO2019,SMRO2020}.
RIXS measurements were performed at the Advanced Photon Source at the 27-ID-B beamline.
The experiments were performed at the \RE{} $L_3$ X-ray absorption
edge ($2p\to 5d$, $E_i=$ 10.532 keV). 
The \BMRO{} experiment utilized the Si(004) reflection while the \SMRO{} and \CMRO{} experiment utilized the Si(440) reflection for high resolution monochromators.
The receiving optics are arranged in the typical Rowland circle geometry to select final photon energy, with a diced, spherically-bent Si(119) crystal analyzer of bending radius 2 m  and a Spectrum Lambda60k detector.
In order to minimize the elastic background from Thomson scattering, the experiments were conducted in a horizontal geometry with $2\theta$ close to $90^\circ$. 
The overall resolution was about 100 meV (FWHM) for the \BMRO{} experiment and about 65 meV for the \SMRO{} and \CMRO{} experiment.
Embedded-cluster quantum chemical computations were carried out using the {\sc molpro} suite of programs, employing the {\sc ewald} program to generate the point-charge embeddings \cite{Molpro,Klintenberg_et_al,Derenzo_et_al}.
The quantum chemical study was initiated as complete active space self-consistent field (CASSCF) calculations with an active orbital space containing the five $5d$ orbitals of the Re ion \cite{olsen_bible,MCSCF_Molpro}.
Post-CASSCF correlation computations were performed at the level of multireference configuration-interaction (MRCI) with single and double excitations out of the Re 5$d$ and O 2$p$ orbitals \cite{olsen_bible,MRCI_Molpro}. 
\par
\begin{figure}[t]
\includegraphics[width=\columnwidth]{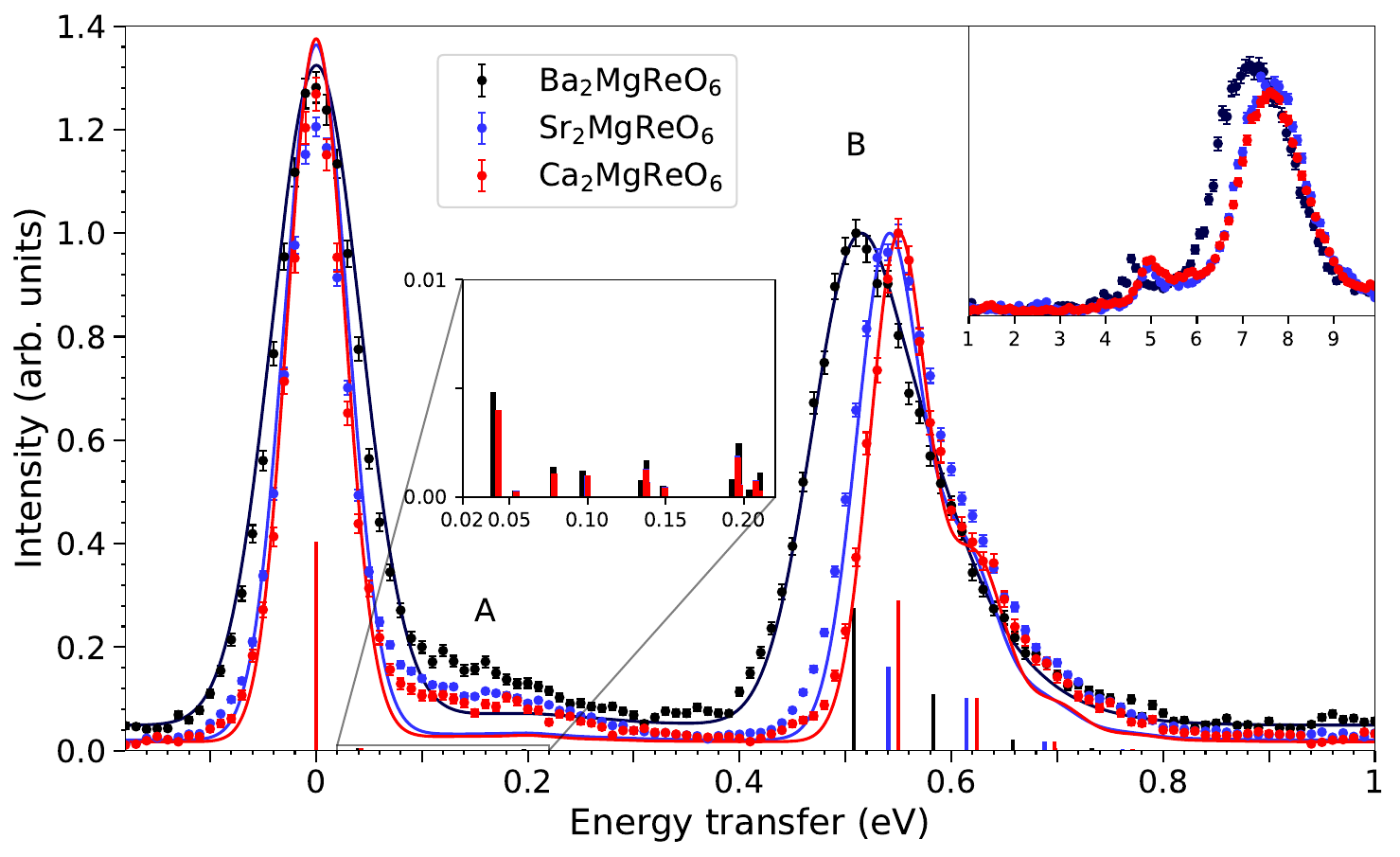}
\caption{Experimental \AMRO{} RIXS spectra up to 1 eV overlaid with vibronic RIXS model. Zoomed inset shows the low energy vibronic modes. All experimental data were obtained at the lowest temperature. Upper right inset shows the RIXS spectra in the higher energy range up to 10 eV.\label{fig:fig2}}
\end{figure}

The RIXS spectra below 1 eV for all three samples show similar features, as shown in Fig. \ref{fig:fig2}.
The higher energy RIXS spectra are shown in the inset of Fig. \ref{fig:fig2} and also show similar features across the series.
The observed features do not show any momentum dependence, see Supplemental Materials \cite{supp} (see also references \cite{Pasztorova2023,Iwahara_RuL3RIXS,Iwahara_IrL3RIXS,Clancy2012,Inui1990,Bersuker1989,BMRO_DFT,AMRO_gap1,AMRO_gap2,SOC_molpro,Figgen_et_al,O-ligands,Fuentealba_1985,Bo_5dRIXS} therein).
A broad feature labelled `A' in Fig. \ref{fig:fig2} is visible as a shoulder to the elastic peak below $\sim$ 300 meV in all compounds.
Its nature is somewhat ambiguous beyond the fact that it corresponds to some intra-$t_{2g}$ excitation.
Possible origins of the feature are discussed at greater length in the supplementary materials.
A sharp, asymmetric feature labelled `B' in Fig. \ref{fig:fig2} is present between 500-550 meV in all compounds, with a systematic energy shift in the peak position to higher energy as the $\mathrm{A_2MgReO_6}$ $\mathrm{A}$ site is substituted from $\mathrm{Ba\to Sr \to Ca}$.
This feature is understood as the \Dii{} transition between the \jgs{} ground state and the \jes{} excited state, see Fig. \ref{fig:fig1} d).
The systematic shift can be understood by examining the effect of static distortion of the \REO{} octahedra on the crystal field levels.
Quantum chemistry calculations summarized in Table \ref{table:table1} estimate that the \REO{} distortions in \SMRO{} and \CMRO{} split the \jgs{} levels by \Diii{} = 50 and 70 meV respectively.
This splitting lowers the ground state energy and correspondingly increases the \Dii{} transition energy by an estimated 30 and 40 meV respectively, in good agreement with the observed shifts of 35 and 45 meV though the calculated transition energies are systematically overestimated, see Table \ref{table:table1}. 
Further details of the compound dependence and comparison with quantum chemistry calculations are discussed in the supplemental materials.\par
\begin{figure}[b]
\includegraphics[width=\columnwidth]{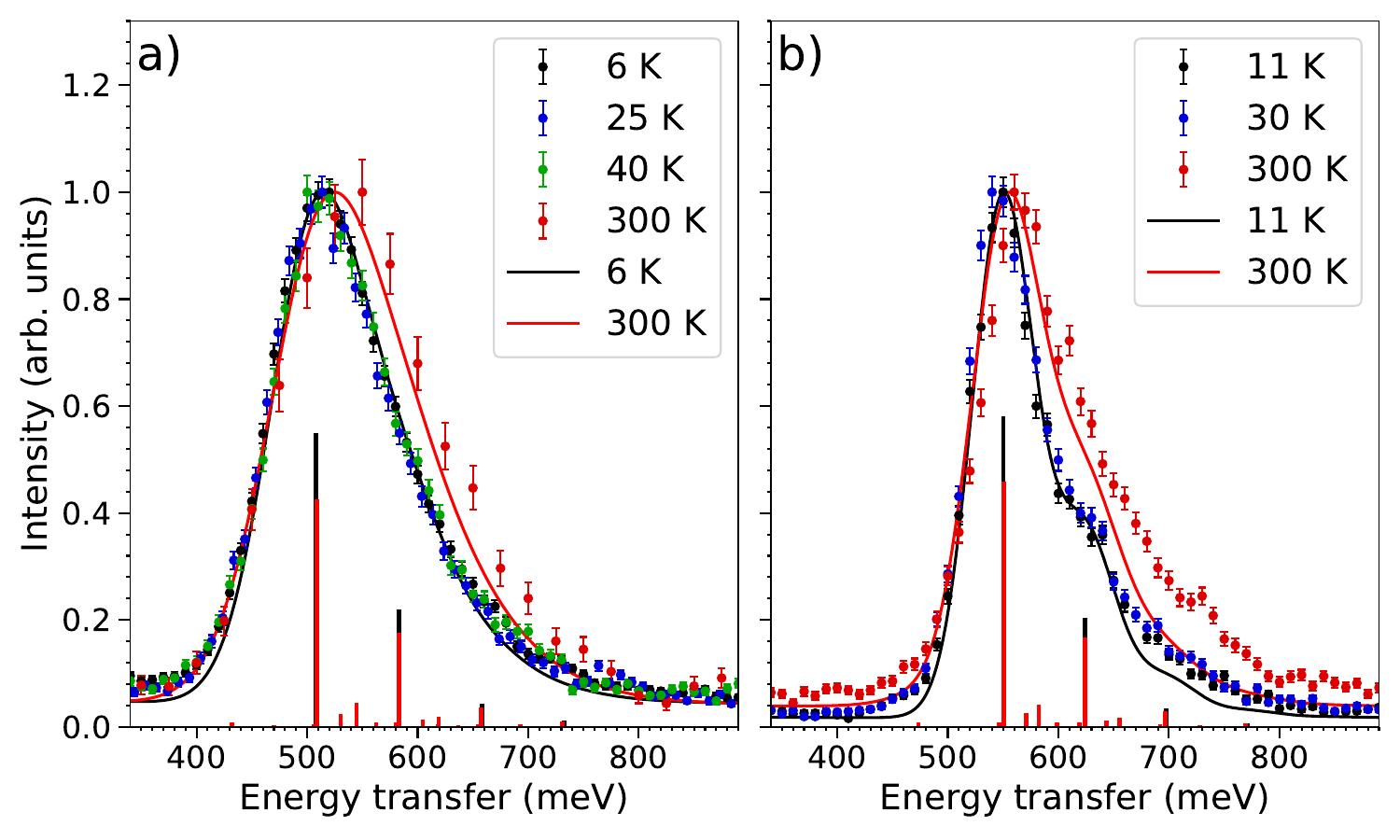}
\caption{a) \BMRO{} and b) \CMRO{} \jes{} feature temperature evolution between base temperature and 300 K overlaid with vibronic RIXS model results.\label{fig:fig3}}
\end{figure}

Temperature dependence of the RIXS spectra below 1 eV is shown in Fig. \ref{fig:fig3} for \BMRO{} and \CMRO{}\footnote{The temperature dependence of the \SMRO{} RIXS spectrum is shown in the supplementary materials.}. 
In the case of \BMRO{} there is no temperature dependence across either $T_m$ or $T_q$ and in \CMRO{} there is no temperature dependence across $T_m$.
On the other hand, temperature dependence is observed over a larger temperature scale when comparing measurements at room temperature and low temperature. 
The RIXS spectra of both \BMRO{} and \CMRO{} show a broadening and shifting towards higher energy of the \jes{} feature at room temperature, as shown in Fig. \ref{fig:fig3}.
This temperature dependence is curious for two reasons.
First, it cannot reflect a change in the non-cubic crystal field splitting of the \jgs{} levels given no structural phase transitions are known to occur between the low and high temperatures.
Second, it is implausible that the shifting reflects a change in SOC based solely on thermal contraction of the lattice; quantum chemistry calculations do not show any change in the SOC using the \BMRO{} \Dii{} transition energy as a proxy, see Table \ref{table:table1}.\par
\begin{table}[t]
\caption{$\mathrm{Re^{6+}}$ $5d^1$ splittings in \AMRO{} at select temperatures, relative energies in eV. Calculated values are obtained from MRCI+SOC.}
\label{table:table1}
\begin{ruledtabular}
\begin{tabular}{r|C{1.75cm}|C{1.75cm}|C{1.75cm}}
&
\textrm{$\mathrm{\Delta_2^{exp.}}$}&
\textrm{$\mathrm{\Delta_3^{calc.}}$}&
\textrm{$\mathrm{\Delta_2^{calc.}}$}\\
\hline
\CMRO{} 300 K& 0.555(5)&0.07 &0.61\\
11 K&0.545(5)&-&-\\
\hline
\SMRO{} 300 K &-& 0.05 & 0.60 \\
11 K&0.535(5)&-&-\\
\hline
\BMRO{} 300 K& 0.52(2) & 0 & 0.57 \\
93 K&-& 0 & 0.57 \\
6 K& 0.50(1)&-&-\\
\end{tabular}
\end{ruledtabular}
\vspace{-5mm}
\end{table} 
While the compound dependent shift is well understood, the asymmetric lineshape of feature B calls for further scrutiny.
We can rule out the asymmetry originating from splitting of the \jes{} level, which is a Kramer's doublet.
With splitting of the \jes{} level ruled out, we next turn to the asymmetry of the feature being owed to the creation of collective magnetic or quadrupolar excitations within the respective ordered phases, i.e. to magnon or quadruplon side-bands.
This is also unlikely given the fact that the asymmetry persists well above all phase transitions.
In contrast, dressing of the spin-orbit excitation with phonon modes and their thermal activation could provide a reasonable explanation for our observations.
Thus, the most likely origin of the \jes{} feature asymmetry and temperature dependence is vibronic coupling to JT active modes, which is supported by our model calculations described below.\par
The model Hamiltonian for a single Re site in an octahedral crystal field consists of SO and dynamic JT terms:
\begin{align}
 \hat{H} &= \hat{H}_\text{SO} + \hat{H}_\text{JT}, 
 \label{Eq:H}
 \\
 \hat{H}_\text{SO} &= \lambda \hat{\bm{l}} \cdot \hat{\bm{s}},
 \label{Eq:HSO}
 \\
 \hat{H}_\text{JT} &= \sum_{\gamma=u,v} \frac{\hslash \omega}{2} \left(\hat{p}_\gamma^2 + \hat{q}_\gamma^2 \right) 
 + \hslash \omega g \left[ 
   \left(-\frac{1}{2}\hat{q}_u + \frac{\sqrt{3}}{2}\hat{q}_v \right) \hat{P}_{yz}
   \right.
 \nonumber\\
 &+ 
 \left.
 \left(-\frac{1}{2}\hat{q}_u - \frac{\sqrt{3}}{2}\hat{q}_v \right) \hat{P}_{zx}
 - \hat{q}_v \hat{P}_{xy}
 \label{Eq:HJT}
 \right].
\end{align}
Here, $\hat{\bm{l}}$ is the $l_{eff}=1$ orbital angular momentum operator for the $t_{2g}$ orbitals \cite{Sugano1970}, 
$\hat{\bm{s}}$ the spin angular momentum operator, 
$\lambda$ the SOC parameter, 
$\hat{q}_\gamma$ the dimensionless normal coordinates for the JT active $E_g$ modes [$\gamma = u$ $(3z^2-r^2), v$ $(x^2-y^2)$], 
$\hat{p}_\gamma$ the conjugate momenta, 
$\hat{P}_\gamma$ ($\gamma = yz, zx, xy$) the projection operator into the $t_{2g}\ \gamma$ orbital, 
$g$ the dimensionless vibronic coupling parameter, 
and $\omega$ the frequency for the JT active mode. 
The zero temperature frequency of the $E_g$ modes is chosen as $\omega_0 = 0.067$ eV with a linear temperature profile $\omega(T)=\omega_0 +cT$ based on the Raman scattering spectra for Ba$_2$MgWO$_6$ \cite{Pasztorova2023}.
$g$ and $\lambda$ are chosen to replicate the experimental data at low temperature; $g$ varies as $1\slash \sqrt{\omega^3}$ with temperature while $\lambda$ is constant based on our MRCI+SOC calculations for \BMRO{}, see Table \ref{table:table1}.
We calculated the cross-section of Re $L_3$-edge RIXS process based on the present theoretical model using the Kramers-Heisenberg formula within the fast collision approximation \cite{Sakurai1967,Luo1993, vanVeenendaal2006}.
To reproduce the shape of RIXS spectrum the RIXS cross-section was convoluted with a Gaussian function matching the resolution of the elastic line.
The vibronic RIXS model is discussed in greater detail in the supplemental materials.\par
The results of model calculations at low temperature are shown in Fig. \ref{fig:fig2} alongside the experimental data. 
The \jes{} feature is well reproduced by values $g$= 1.325, 1.275, 1.25 and $\lambda$ = 0.311, 0.337, 0.343 eV for A = Ba, Sr, Ca respectively.
We note that the values of $\lambda$ are unequal to compensate for the increase in the \jes{} feature energy due to \REO{} distortion not included in the pure octahedral symmetry of the model.
In order to estimate the energy scale of vibronic coupling we examine the modelled contributions to the \jgs{} $\to1/2$ transition energy.
As previously discussed, the dynamic JT effect lowers the ground state energy and, consequently, increases \Dii{} from the usual $3\lambda/2$ by an amount that grows monotonically with the vibronic coupling strength.
To best estimate the contribution from vibronic coupling, then, we inspect the nearly undistorted case of \BMRO{} where \Diii$\sim 0$ and \Dii{}$\sim 3\lambda/2$ in the absence of the dynamic JT effect.
In this case, the model value of $\lambda$ gives $3\lambda/2$= 0.467 eV which, when compared to the experimental \jgs{} $\to 1/2$ transition energy of 0.5 eV, indicates that the transition energy increases by $\sim$ 0.033 eV due to the dynamic JT effect.
Ignoring static distortions, this same subtraction gives similar dynamic JT effect contributions of $\sim$ .03 eV in \SMRO{} and \CMRO{}.
Thus, we conclude that the energy scale of vibronic coupling is one order of magnitude smaller than that of spin-orbit coupling in \AMRO{}.
With regards to the increase in $g$ from $\mathrm{Ca\to Sr \to Ba}$, one possibility is that this reflects the corresponding decrease in static JT distortion across the series by virtue of the fact that large vibronic coupling strengths can suppress static distortions in $5d^1$ DPs.
Another possibility is that it reflects differences in the energy of the JT active phonon modes between compounds.
This viewpoint is supported by a comparison of the room temperature JT active phonon energies of Ba$_2$MgWO$_6$ and Ca$_2$MgWO$_6$, the latter of which is $\sim 7\%$ more energetic (0.072 eV compared to 0.067 eV) \cite{Ca2MgWO6_Raman}.
Such an increase in the phonon energy would correspond to a decrease of $\sim 10\%$ in $g$, more than sufficient to effectuate the observed decrease of $\sim 6\%$.
It is important to note, however, that the worse resolution of the \BMRO{} experiment as well as the fact that our model does not account for the impact of static JT distortions may also have impacts on the estimated $g$ parameters.\par
The temperature evolution of the vibronic RIXS model is shown in Fig. \ref{fig:fig3}, from which we see that our model can qualitatively capture a shift to higher energy and broadening of the high energy tail but quantitatively underestimates both aspects.
The peak shift and broadening can be matched by arbitrarily increasing $g$ at high temperature, however, the physical mechanism for such an increase is not understood and as such is not done.
Of the possible mechanisms that may affect $g$, we can rule out the effect of phase transitions in \BMRO{} and \CMRO{} given the temperature independence of the RIXS spectra across these.
On the other hand, because $g\propto 1/\sqrt{\omega^3}$, it is plausible that the missing enhancement of $g$ comes from underestimation of the \AMROx{} phonon energy temperature dependence due to anharmonic effects \cite{BFRO_phonon}.
Future efforts to characterize the temperature dependence of the JT active phonon energies in the \AMRO{} compounds would help to resolve this question and point to the nature of the changes in $g$ between them.\par

In summary, we have studied the family of \AMRO{} compounds using \RE{} $L_3$-edge RIXS. 
Our results show a splitting of the \RE{} $t_{2g}$ levels into the expected \jgs{} and \jes{} levels.
The splitting is found to be in the range of 0.50-0.55 eV for the \AMROx{} family, with a systematic shift upwards from $\mathrm{Ba\to Sr \to Ca}$ indicative of the distortion of the \REO{} octahedra in the respective compounds.
The \jes{} feature is observed to be distinctly asymmetric and broadens and shifts to higher energy at high temperature.
These properties are well explained by dressing of the \jes{} feature with lattice vibrations, implying the existence of significant vibronic coupling in the \AMROx{} compounds and furthermore indicating a spin-orbit-lattice entangled ground state. 
This takeaway is reinforced by the construction of a vibronically coupled RIXS model which is shown to be in good agreement with both the lineshape and temperature evolution of the observed spectra.
One distinct consequence of a spin-orbit-lattice entangled ground state is that the SOC strength $\lambda$ cannot directly be extracted, necessitating the inclusion of vibronic coupling to the typical crystal field + SOC picture when attempting to untangle the $5d$ level splittings.

\begin{acknowledgments}
Work at the University of Toronto was supported by the Natural Sciences and Engineering Research Council (NSERC) of Canada through the Discovery Grant No. RGPIN-2019-06449 and the Alliance International Catalyst Quantum Grant No. ALLRP 576422 - 22, Canada Foundation for Innovation, and Ontario Research Fund.
This work was partly supported by Grant-in-Aid for Scientific Research (Grant No. 22K03507 and 20H01858) from the Japan Society for the Promotion of Science, JSPS Bilateral Program (Grant No. JPJSBP120239915) and the Iketani Science and Technology Foundation.
P.\,B. and L.\,H. acknowledge financial support from the German Research Foundation (Deutsche Forschungsgemeinschaft, DFG), project 441216021, and technical assistance from U.~Nitzsche.

\end{acknowledgments}
\bibliography{refs}

\end{document}